\documentclass[12pt,a4paper]{article}
\usepackage[]{amsmath,amssymb}
\usepackage{graphics,epsfig}
\usepackage{graphicx, color}
\usepackage{geometry}

\begin{document}

\date{}
\title{Entanglement entropy for non-coplanar regions in quantum field theory}
\author{David D. Blanco\footnote{e-mail: blancod@ib.cnea.gov.ar} 
 \, and Horacio Casini\footnote{e-mail: casini@cab.cnea.gov.ar} \\
{\sl Centro At\'omico Bariloche,
8400-S.C. de Bariloche, R\'{\i}o Negro, Argentina}}
\maketitle

\begin{abstract}
We study the entanglement entropy in a relativistic quantum field theory for regions which are not included in a single spatial hyperplane. This geometric configuration cannot be treated with the Euclidean time method and the replica trick. Instead, we use a real time method to calculate the entropy for a massive free Dirac field in two dimensions in some approximations. We find some specifically relativistic features of the entropy. First, there is a large enhancement of entanglement due to boosts. As a result, the mutual information between relatively boosted regions does not vanish in the limit of zero volume and large relative boost. We also find extensivity of the information in a deeply Lorentzian regime with large violations of the triangle inequalities for the distances. This last effect is relevant to an interpretation of the amount of entropy enclosed in the Hawking radiation emitted by a black hole. 
\end{abstract}

\section{Introduction}
In the context of quantum field theory (QFT) the term "entanglement entropy" usually refers to the entropy $S(V)$ of the vacuum state reduced to a region $V$ of the space. It essentially measures the entropy contained in the vacuum fluctuations in this region. The interest in this subject arose initially from an interpretation of the black hole entropy in terms of entanglement entropy \cite{bh}. Later, several different applications have been developed, showing that the entanglement entropy of the vacuum contains information on important aspects of the QFT, including renormalization flow \cite{cteo,rfl}, topological order \cite{top}, phase transitions \cite{pt} and confinement \cite{con}. Recently, the possibility of using the entanglement entropy as a tool for developing the AdS-CFT dictionary has attracted much interest \cite{ryu}. 

\begin{figure}
\centering
\leavevmode
\epsfysize=5cm
\epsfbox{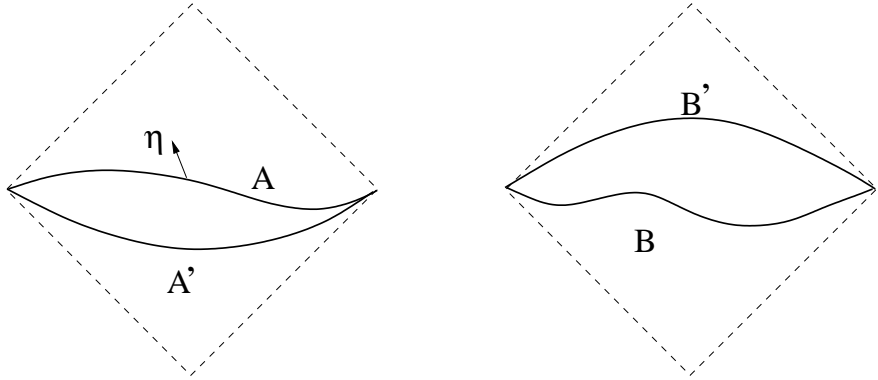}
\bigskip
\caption{The entanglement entropy is a function of pieces of spatial surfaces like the set $A$ in this figure (light like lines are shown at $\pm 45^o$). Spatial sets having the same causal domain of dependence (diamond shaped set) such as $A$ and $A^\prime$, have the same reduced density matrix. In consequence, a regularization independent quantity such as the mutual information $I(A,B)$ coincides with the one between equivalent regions $I(A^\prime,B^\prime)$.  }
\label{paralelas}
\end{figure}

In contrast with the more intuitive idea of the entropy of a substance contained in a box, which under normal circumstances persists on time, the entanglement entropy of a region has to be defined for a fixed instant of time (see figure (\ref{paralelas})). After this moment, the reduced state on the region $V$ will spread out at the velocity of light. In order to impede this spreading, some material box  imposing boundary conditions would be necessary. But this is not what we are willing to do, since this further element, the boundary condition, spoils the very nature of the entanglement entropy, of being a quantity depending {\sl only} on the geometry of $V$ and the particular QFT.   

Therefore entanglement entropy has to be thought as a quantity localized in space-time \cite{geo}. In the relativistic case, $V$ can be any $d$-dimensional spatial region (in a $d+1$-dimensional space-time), included as a part of a Cauchy surface, but not necessarily contained in a flat spatial hyperplane. This space-time nature of entanglement entropy is fundamental to the entropic c-theorem in two dimensions \cite{cteo}, but has otherwise not received much attention in the literature\footnote{However, effects of boosts on the entanglement of the spin degrees of freedom of relativistic particles have been intensely studied. See for example the review papers \cite{uni}.}. 

This may be attributed to the fact that if $V$ is not contained in a single spatial hyperplane the usual Euclidean method to compute entanglement entropy based on the replica trick becomes inapplicable in a direct way. A $d$-dimensional surface in Euclidean space corresponds to a $d$-dimensional spatial surface in Minkowski only if this is a flat surface. Otherwise, the Minkowskian result should follow from the Euclidean one through a complicated analytic continuation in the space of regions.    

In this paper we analyze the behavior of the entanglement entropy for non coplanar regions in the simplest QFT model given by a free fermion in two dimensions. We choose a fermion field instead of a scalar one, since this later involves the treatment of more singular kernels \cite{review}, and in addition, in the two dimensional case it develops an infrared divergence in the massless limit. We use a real time method based on the explicit expression of the reduced density matrix for the free case \cite{pes,fermion1}. The entanglement entropy for several disjoint intervals lying on a single spatial line was calculated in  \cite{fermion1} using this approach and a small mass expansion. We extend these techniques here to obtain the general result for non coplanar sets involving relatively boosted intervals.   

The entanglement entropy in QFT contains divergences which are proportional to local terms on the boundary of $V$. The regularization independent information can be isolated using the mutual information function\footnote{Given two non-overlapping regions several well defined regularization independent "entropic" quantities and entanglement measures can be constructed \cite{narn}.} 
\begin{equation}
I(A,B)=S(A)+S(B)-S(AB)
\end{equation}
for two disjoint regions $A$ and $B$. This quantity has the interpretation of the amount of shared information between these regions, and is always positive. It is also increasing with the size of $A$ and $B$, and it gives an upper bound on correlations between these two sets. We calculate this function in some specific approximation, and find that this shared information can be greatly increased by the relative boosts between $A$ and $B$. Surprisingly, the boost enhancement is such that the mutual information remains no zero for zero volume sets, provided they lie on null surfaces. 

An important physical problem which naturally involves the entanglement between regions with a large relative boost is the localization of information in the Hawking radiation process, in the semiclassical regime. Indeed, the entropy in the Hawking radiation is entanglement entropy with the region hidden across the horizon. The entropy in the Hawking radiation contained in a finite region of space is rather small and cannot be isolated from the area terms in the entanglement entropy, which are regularization dependent \cite{bhh}. This can be overcome using a regularization independent measure of information. A natural choice is $I(A,B)$, being $A$ the black hole (or some piece of it near the bifurcation surface) and $B$ a region far outside the black hole, containing Hawking radiation. These two regions are related by an exponentially large redshift, which arguably may be simulated by a large relative boost in flat space. 
 
Physical intuition dictates that the information in the radiation region should be spatially extensive, at least for regions which are far from the black hole and are large with respect to the typical radiation wavelength.  
This property of extensivity is however non mathematically guaranteed because mutual information is in general a non extensive quantity  \cite{bhh,extensiva}. The deviation from extensivity is measured by the tripartite information
\begin{equation}
I(A,B,C)=I(B,A,C)=I(A,C,B)=I(A,B)+I(A,C)-I(A,BC)\,,
\end{equation} 
which is zero in the extensive case $I(A,BC)=I(A,B)+I(A,C)$. Thus, an important question related to the entanglement entropy for non coplanar surfaces is whether an extreme Lorentzian geometry may guide us to find a general principle making the mutual information extensive. 

 We study this question with our toy QFT model. The massless fermion in two dimensions has extensive mutual information, but this property fails for massive fields \cite{fermion1}. In the black hole interpretation, the presence of a mass in two dimensions connects the ingoing and outgoing modes, and simulates the backscattering which occurs for massless fields in more than two dimensions.  We find here that some special configurations with high relative boosts between the different components can restore extensivity in the massive case. We relate this property to large violations of the triangle inequalities for the involved distances, which are allowed by the Lorentzian geometry. These large violations of the triangle inequalities are also a key ingredient of the black hole evaporation geometry.

\section{Entanglement entropy for a Dirac fermion}
  In a two dimensional space-time $V$ is a spacelike curve, which can have several connected components. We are interested in a free Dirac field $\Psi(x)$, and write $\Psi(s)$ for the field $\Psi(x(s))$, where $x(s)$ is a parameterization of the points of $V$ by the distance parameter $s$ along the curve.  

The general expression of the reduced density matrix $\rho_V$, corresponding to the global vacuum state, and the region $V$, was obtained in \cite{fermion1} in terms of the field correlator. We will not need this expression here, but only the corresponding one to the associated entanglement entropy $S(V)$. This is given by \cite{fermion1}
\begin{equation}
S(V)=-\textrm{tr}\left[(1-\tilde{C})\log(1-\tilde{C})+\tilde{C}\log(\tilde{C})\right]\,,\label{pelada}
\end{equation}
where $\tilde{C}$ is the operator with kernel 
\begin{equation}
\tilde{C}(s_1,s_2)=\left\langle0 \right|\Psi(s_1)\,\Psi^\dagger(s_2) \left|0\right\rangle\bar{\eta}(s_2)\,.\label{tree}
\end{equation} 
 Here
 $\bar{\eta}(s)=\gamma^0 \gamma^\mu \eta_\mu(s)$,
and $\eta_\mu (s)$ is the future directed unit vector normal to the curve $V$ at the point $x(s)$.  We have $\bar{\eta}(s)=\bar{\eta}(s)^\dagger$ is hermitian and positive, with $\det{\bar{\eta}}=1$, $\bar{\eta}^{-1}=\gamma^0 \bar{\eta}\gamma^0=\gamma^\mu \eta_\mu(s)\gamma^0$. We are using a $(1,-1)$ signature for the metric (time-like vectors have positive square).

This result, as well as the explicit expression for the density matrix, follows from the requirement
 $\textrm{tr}( \rho_V {\cal O})=\langle 0|{\cal O}|0\rangle$  for any operator ${\cal O}$ localized in $V$,
    together with the very simple structure of the vacuum expectation value for the field polynomials in the free case \cite{review,pes}. The density matrix and the entropy require regularization. A more rigorous treatment based on the KMS condition 
 is in \cite{reh}.

The density matrix represents the vacuum state on the whole algebra of operators localized in $V$. Because of causality, this coincides with the algebra of operators localized in any other spatial surface having the same domain of dependence as $V$ \cite{fermion1}. For physical, finite quantities, such as the mutual information $I(A,B)$ between two regions, this means that $I(A,B)=I(A^\prime,B^\prime)$ for any pairs of surfaces $A,$ $A^\prime $ and $B$, $B^\prime$, having the same domain of dependence. This is illustrated in figure (\ref{paralelas}).

In two dimensions we have for the field correlator
\begin{eqnarray}
\left\langle0 \right|\Psi(x)\,\Psi^\dagger(y) \left|0\right\rangle &=&\frac{1}{4} \partial_\mu \left[\theta{\left((x-y)^2\right)}\textrm{sgn}(x^0-y^0)\right] \, \gamma^\mu\gamma^0  \nonumber\\
&+&\frac{m}{2\pi} K_0(m|x-y|) \gamma^0  -\frac{i m}{2 \pi} \, K_1(m |x-y|) \frac{(x-y)_\mu}{|x-y|}\gamma^\mu \gamma^0 \, \,,\label{yuyi}
\end{eqnarray}
where $K_a(x)$ is the standard modified Bessel function, $\gamma^\mu$ are the Dirac matrices, and $\vert x\vert=\sqrt{-x_\mu x^\mu}$. The contribution of the first term on the right hand side of (\ref{yuyi}) to the kernel $\tilde{C}$ of eq. (\ref{tree}) is just half the identity kernel, $1/2 \,\delta(s_1-s_2)$, for any curve $V$.

In order to do perturbations it is convenient to express (\ref{pelada}) in terms of the resolvent kernel $\tilde{R}=(\tilde{C}-1/2+\beta)^{-1}$. This is
\begin{equation}
S(V)=-\int^\infty_{1/2} d\beta\, \textrm{tr}\left[\left(\beta-1/2\right) \left(\tilde{R}(\beta)-\tilde{R}(-\beta)\right)-\frac{2\beta}{\beta+1/2}\right]\,.\label{fas}
\end{equation}

\section{The small mass expansion}  
According to (\ref{pelada}) the evaluation of the entropy requires the resolution of the kernel (\ref{yuyi}) involving Bessel functions. Unfortunately this is not known at present. Here we take advantage of the known expressions for the spectral resolution of the massless kernel in order to make a small mass expansion for the entropy.  This means that we have to consider all the typical distances in $V$ to be smaller than $m^{-1}$. It is assumed this holds for the rest of the paper, unless otherwise noted.

The expansion for the correlator reads $C=C_0+C_1+C_2+...$, with 
\begin{eqnarray}
C_0 (x,y)&=&\frac{1}{4} \partial_\mu \left[\theta{\left((x-y)^2\right)}\textrm{sgn}(x^0-y^0)\right] \, \gamma^\mu\gamma^0 \, 
 - \frac{i}{2\pi}\frac{(x-y)_\mu}{|x-y|^2}\gamma^\mu \gamma^0\,, \label{horacio}\\
C_1 (x,y)&=&-\frac{m}{2\pi}\left(\gamma_E+\log\left(\frac{m|x-y|}{2}\right)\right) \gamma^0 \,,\\
C_2 (x,y)&=&-\frac{i m^2}{4\pi} \left(\gamma_E -\frac{1}{2} +\log\left(\frac{m|x-y|}{2}\right) \right) (x-y)_\mu\gamma^\mu\gamma^0\,,
\end{eqnarray}
and $\gamma_E$ is the Euler constant. 
The perturbative expansion in the mass involves non-commuting kernels $C_0$ and $C_i$, $i>0$. Hence, it is convenient to use the formula for the entropy in terms of the resolvent, and expand the resolvent as
\begin{equation}
\tilde{R}_V(\beta)=\tilde{R}_V^0(\beta)-\tilde{R}_V^0(\beta)\tilde{C}_1 \tilde{R}_V^0(\beta)-\tilde{R}_V^0(\beta)\tilde{C}_2 \tilde{R}_V^0(\beta)+\tilde{R}_V^0(\beta)\tilde{C}_1 \tilde{R}_V^0(\beta)\tilde{C}_1 \tilde{R}_V^0(\beta)-...\,,
\label{financiar}
\end{equation}  
where $\tilde{C}_i(x,y)=C_i(x,y)\bar{\eta}(y)$, and $\tilde{R}_0=(\tilde{C}_0-1/2+\beta)^{-1}$.
A straightforward analysis of the possible terms for the expansion in entropy formula shows that the series can be written in terms of powers of $m$ and $\log(m)$ as  \cite{fermion1,fermion2}
 \begin{equation}
 S=\sum_{i=0}^\infty \sum_{j=0}^{2 i}S_{i,j}=\sum_{i=0}^\infty \sum_{j=0}^{2 i} s_{i,j}\,\, m^{2 i}\log^j(m)\,.
 \end{equation}
Massive corrections for the entanglement entropy of free fields in more dimensions are studied in \cite{wi}. 

\subsection{The massless contribution}
In the massless case the problem factorizes in the two chiralities. The massless correlator $\tilde{C}_0$ diagonalizes in the chiral representation for the spinors, that is, the base where $\gamma^3=\gamma^0 \gamma^1=\textrm{diag}(1,-1)$ is diagonal. In order to see this we parameterize the coordinate differential tangent to the curve $V$  as 
\begin{equation}
(dx^{\mu })\equiv\left( \sinh \alpha(s) ,\cosh \alpha(s) \right) ds\,.
\end{equation}
Hence, the unit vector normal to $V$ is
\begin{equation}
(\eta _{\mu })\equiv\left( \cosh \alpha ,-\sinh \alpha \right)\,, 
\end{equation}
and we have in the chiral representation for the Dirac matrices
\begin{equation}
\bar{\eta}=
\begin{pmatrix}
e^{-\alpha } & 0 \\ 
0 & e^{\alpha }
\end{pmatrix}\,.\label{etaeta}
\end{equation}
Using null coordinates
\begin{equation}
u_\pm=x^0\pm x^1\,,\label{kad}
\end{equation}
we can write $\tilde{C}_0$ as  
\begin{equation}
\tilde{C}_0\equiv\begin{pmatrix}
D_-(u_-^x , u_-^y) & 0 \\ 
0 & D_+(u_+^x , u_+^y)
\end{pmatrix}\,.\label{tuyoa}
\end{equation}
Here $D_\pm$ are scalar kernels having the same expression
\begin{equation}
D_\pm(u_1,u_2)=\frac{1}{2} \delta(u_1-u_2) - \frac{i}{2 \pi} \,  \frac{1}{(u_1-u_2)}\,,\label{tuyo}
\end{equation}
but different domains, given by the projections $V_\pm$ of $V$ on the null axis. The kernel (\ref{tuyo}) is understood in the principal value regularization.
Note that we have used the relations 
\begin{equation}
du_{+}=e^{\alpha }ds\,,\hspace{2cm} du_{-}=-e^{-\alpha }ds\,,\label{ferro}
\end{equation}
 in order to change coordinates and rewrite the operator $\tilde{C}_0$. In (\ref{tree}) it is expressed as a kernel in the distance variable, while in (\ref{tuyoa}) it is understood that $D_+(u_+^x , u_+^y)$ and $D_-(u_-^x , u_-^y)$ act as kernels on the $u_+$ and $u_-$ variables respectively.

\begin{figure}
\centering
\leavevmode
\epsfysize=8cm
\epsfbox{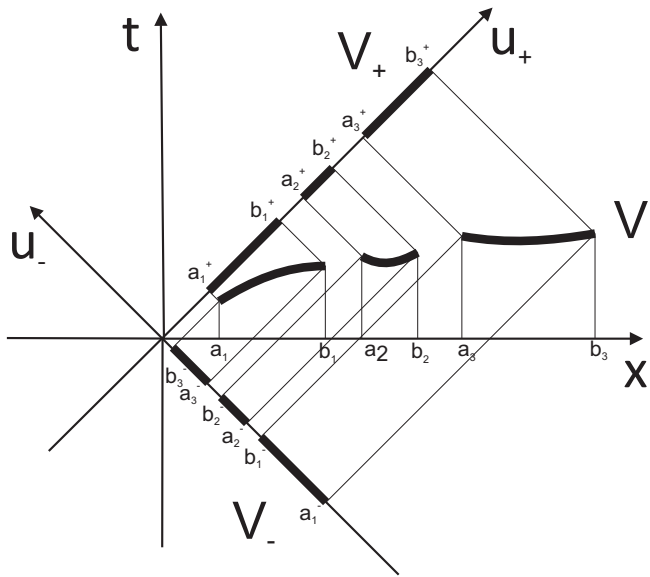}
\bigskip
\caption{The projections $V^\pm$ of the set $V$ on the null coordinate axis. $V$ has three connected components in this example. The labeling of the extreme points of the intervals in $V^\pm$ is done in increasing order for the null coordinates. Note for example that $b_3^+$ and $a_1^-$ are null coordinates corresponding to the same point.}
\label{fy}
\end{figure}

Let us call $l_1,...,l_n$ to the left extreme points of the $n$ different connected components of $V$ ordered from left to right in the spatial coordinate, and $r_1,...,r_n$ for the right extreme points. The projection of $V$ onto the null coordinates is formed by a union of disjoint intervals, which we call $V_\pm=(a^\pm_1, b^\pm_1)\cup ...\cup (a^\pm_n,b^\pm_n)$,with $a_i^\pm< a_{i+1}^\pm$. 
Hence we have simply $a_i^+=l_i^+$ and $b_i^+=r_i^+$, while $a_i^-=r_{n-i+1}^-$ and $b_i^-=l_{n-i+1}^-$ (see figure (\ref{fy})). The ordering for the extreme points in $V^-$ is inverted with respect to the natural left to right ordering because of the sign in (\ref{kad}).

The spectral decomposition of the scalar kernels $D_\pm$ in arbitrary multi-component sets is known \cite{fermion1,mush}. We review the main properties of the kernel $(x-y)^{-1}$ in the Appendix. Using this decomposition, a straightforward calculation of eq. (\ref{pelada}), which we do not repeat here, leads to the result  \cite{fermion1}  
\begin{eqnarray}
S_{0,0}&=&S^+_{0,0}+S^-_{0,0}\,,\\
S_{0,0}^\pm&=&\frac{1}{6} \left( \sum_{i,j}\log|b^\pm_i-a^\pm_j|-\sum_{i<j} \log|a^\pm_i-a^\pm_j| -\sum_{i<j} \log|b^\pm_i-b^\pm_j|-n \log \epsilon \right)\,.\label{sesen}
\end{eqnarray}
Here $\epsilon$ is a short distance cutoff. It appears in the calculation of the trace in (\ref{pelada}), where the integration is taken from $a_i^\pm+\epsilon$ to $b_i^\pm-\epsilon$ in each interval. For a single interval the entropy is proportional to the logarithm of the interval length, and this is a general result for two-dimensional conformal theories \cite{hol}.

The mutual information is of course finite. A simple calculation shows it can be written as \cite{extensiva}
 \begin{equation}
 I_{0,0}(A,B)=\frac{1}{3}\int_A \int_B ds_A\, ds_B\, \frac{\eta_A^\mu \,\eta_B^\nu}{\vert x^A-x^B\vert^2} \left(g_{\mu\nu}-2\frac{(x_\mu^A-x_\mu^B)(x_\nu^A-x_\nu^B)}{\vert x^A-x^B\vert^2}\right)\,, \label{quar}
 \end{equation}  
where $x^A$ and $\eta_A$ (repectively $x^B$ and $\eta_B$) are functions of $s_A$ ($s_B$) giving a parameterization of the position and the normal along the curve $A$ ($B$). It can also be written as a sum over the two chiralities using (\ref{sesen}). In particular for two single interval sets $A$ and $B$ we have 
\begin{equation}
I_{0,0}(A,B)=\frac{1}{6}\log\left(\frac{(b^+_2-b^+_1)(a^+_2-a^+_1)}{(a^+_2-b^+_1)(b^+_2-a^+_1)}\right)+\frac{1}{6}\log\left(\frac{(b^-_2-b^-_1)(a^-_2-a^-_1)}{(a^-_2-b^-_1)(b^-_2-a^-_1)}\right)\,,\label{miro}
\end{equation}
which is a function of the cross-ratios of the projections of the extreme points on the null axis. For more general CFT the equation corresponding to (\ref{miro}) may contain an additional term which is function of the cross ratios \cite{cc1}. However, unlike the logarithms in (\ref{miro}) this new term is a  bounded function \cite{cteo}. 
 
The expression (\ref{quar}) as a double integral shows immediately that the mutual information is extensive in the massless limit,
\begin{equation}
I_{0,0}(A,B,C)=0.
\end{equation}

In the following section we compute the first terms in the small mass expansion of the entropy. It  is then natural to express all the kernels in the chiral base, where the massless one is diagonal. 
In order to do this we will need to write traces of products of operators in the chiral representation for the Dirac matrices, and as integrals of kernels acting on the null coordinates. This is done with the useful formula
\begin{eqnarray}\label{teoremagen}
&&\int ds_{1}ds_{2}...ds_{n}\textrm{tr}\left[ \tilde{A}^{1}\left(
s_{1},s_{2}\right) \tilde{A}^{2}\left( s_{2},s_{3}\right) ...\tilde{A}
^{n}\left( s_{n},s_{1}\right) \right] = \\
&&\underset{q_{1},q_{2},...,q_{n}=+,-}{\sum }\,\,\,\,\underset{V_{q_{1}}}{\int }
du_{q_{1}}\underset{V_{q_{2}}}{\int }du_{q_{2}}...\underset{V_{q_{n}}}{
\int }du_{q_{n}}A_{q_{1}q_{2}}^{1}\left( u_{q_{1}},u_{q_{2}}\right)
A_{q_{2}q_{3}}^{2}\left( u_{q_{2}},u_{q_{3}}\right)
...A_{q_{n}q_{1}}^{n}\left( u_{q_{n}},u_{q_{1}}\right)\nonumber\,.
\end{eqnarray}
In this formula $\tilde{A}^i(s_1,s_2)=A^i(s_1,s_2)\bar{\eta}(s_2)$ is a kernel in distance coordinates, and $A^i_{p q}(u_p,u_q)$, with $p,q=\pm$, is the $(p,q)$ matrix element of $A^i$ (without the $\bar{\eta}$ factor), in the chiral representation, evaluated at the points $s_1(u_p)$ and $s_2(u_q)$. The eq. (\ref{teoremagen}) follows directly from (\ref{etaeta}) and (\ref{ferro}). 

\subsection{Massive corrections}

In this section we calculate the leading log terms $S_{2,2}$ and $S_{2,1}$ in the expansion for the entropy, using the expansion for the resolvent (\ref{financiar}). We need to expand to this order because $S_{2,1}$ is the leading non extensive term. The massless resolvent is diagonal
\begin{equation}
R^0_V(\beta)=
\begin{pmatrix}
R^0_{V^-}(\beta) & 0 \\ 
0 & R^0_{V^+}(\beta)
\end{pmatrix}\,.
\end{equation}
Using the spectral decomposition of the Appendix we have 
\begin{equation}
R_{V^\pm}^0(\beta)(u^1_\pm,u^2_\pm)=\sum_{k=1}^{n}\int^{+\infty}_{-\infty} ds\, \Psi_s^{k\pm}(u^1_\pm) \, m(\beta , s)\,   \Psi_s^{k\pm\,*}(u_\pm^2)\,,\label{tisei}
\end{equation} 
with
\begin{equation}
m(\beta , s)=\left(\beta -\frac{1}{2}\tanh(\pi s)\right)^{-1}\,.
\end{equation}
The leading massive term in the series is proportional to $m^2 \log^2(m)$ and corresponds to the fourth term in (\ref{financiar}). We display with certain detail the calculation of this term in order to exemplify the main technical steps. These are essentially the same for the calculation of the following terms in the expansion. We have 
\begin{eqnarray}
 S_{2,2}&=&-\frac{m^{2}}{4\pi ^{2}}\log ^{2}(m)\overset{+\infty }{
\underset{1/2}{\int }}d\beta \left( \beta -1/2\right) \textrm{tr}\left[
\tilde{R}_{V}^{0}(\beta ){\bf 1}\,\gamma ^{0}\bar{\eta}\tilde{R}_{V}^{0}(\beta ){\bf 1}\,\gamma ^{0}\bar{\eta}\tilde{R}_{V}^{0}(\beta
)\right.\nonumber\\
&&\hspace{6cm}\left.-\tilde{R}_{V}^{0}(-\beta ){\bf 1}\,\gamma ^{0}\bar{\eta}\tilde{R}_{V}^{0}(-\beta ){\bf 1}\,\gamma ^{0}\bar{\eta}\tilde{R}_{V}^{0}(-\beta
)\right]\,,\label{esedodo}
\end{eqnarray}
where the kernel ${\bf 1}(x,y)=1$ for all $x$, $y$. 
Using the representation (\ref{teoremagen}) the first term in the trace in this last equation writes (the second one follows just replacing ($\beta\rightarrow -\beta$))
\begin{equation}
\begin{array}{c}
\underset{V^{-}}{\int }dx\underset{V^{-}}{\int }dy\underset{V^{+}}{\int }du
\underset{V^{+}}{\int }dwR_{V^{-}}^{0}(\beta )^{2}(x,y)R_{V^{+}}^{0}(\beta
)(u,w)+\\+\underset{V^{+}}{\int }dx\underset{V^{+}}{\int }dy\underset{V^{-}}{
\int }du\underset{V^{-}}{\int }dwR_{V^{+}}^{0}(\beta
)^{2}(x,y)R_{V^{-}}^{0}(\beta )(u,w)\,.
\end{array}
\end{equation}
This can be further expanded using the spectral decomposition as
\begin{equation}
\overset{n}{\underset{k,k^{\prime }=1}{\sum }}\,\,\underset{-\infty }{\overset{+\infty }{\int }}
ds\,ds^{\prime }\,m(\beta ,s)^2m(\beta ,s^{\prime })\left\vert \Lambda _{-}(s,k)\right\vert ^{2}\left\vert
\Lambda _{+}(s^\prime,k^\prime)\right\vert ^{2}+(+\leftrightarrow -)\,.
\label{large}
\end{equation}
Here we have written
\begin{equation}
\Lambda_{\pm}(s,k)=\int_{V_\pm}  dx\,  \Psi_s^{k\pm}(x)\,.
\end{equation}
Combining eqs. (\ref{nos}) and (\ref{nuno}) of the Appendix we have the useful property
\begin{equation}
\sum_{k=1}^n |\Lambda_\pm(s,k)|^2=\frac{\pi}{2} \textrm{sech}^2(\pi s) L_\pm\,,\label{yuy}
\end{equation}
with $L_\pm=\sum_i (b^\pm_i-a^\pm_i)$.
Using (\ref{yuy}), the sums in (\ref{large}) are solved, and we end up with integrals in $s$, $s^\prime$ and $\beta$ in (\ref{esedodo}). These can be evaluated analytically, leading to 
\begin{equation}
 S_{\,2,2}=-\frac{m^2}{6}  \log^2(m) \,L_+ \,L_-\,.\label{leading}
\end{equation}

This simple quadratic expression also gives an extensive contribution, 
\begin{equation}
I_{2,2}(A,B,C)=0\,.
\end{equation}
In order to find the first non-extensive term we calculate the next logarithmic order proportional to $m^2 \log (m)$. This is produced from the third and fourth terms in (\ref{financiar}) (the second term does not contribute). Following \cite{fermion1}, we write this contribution as a sum of three terms,
\begin{equation}
 S_{\,2,1}=s_{2,1}\,\,m^2 \log(m)=\Delta_1+\Delta_2+\Delta_3\,.
\end{equation}
 We call $\Delta_1$ to the contribution coming from the terms in $C_1$ which does not contain $\log|x-y|$, $\Delta_2$ to the one coming from $C_2$, and $\Delta_3$ to the one involving the term proportional to  $\log|x-y|$ in $C_1$. More explicitly, 
\begin{eqnarray}\label{delta1cuenta}
\Delta_{1}&=& -\frac{m^{2}}{2\pi ^{2}}\left( \Gamma _{E}-\ln
2\right) \log (m)\overset{+\infty }{\underset{1/2}{\int }}d\beta \left( \beta
-1/2\right) \textrm{tr}\left[  \tilde{R}^{0\, 2}{\bf 1} \gamma ^{0}\bar{\eta}
\tilde{R}^{0}{\bf 1}\gamma ^{0}\bar{\eta}-\left( \beta \leftrightarrow -\beta
\right) \right]\,,
\\
\label{delta2cuenta}
\Delta_{2}&=& -\frac{im^{2}}{4\pi }\log (m)\overset{+\infty }{
\underset{1/2}{\int }}d\beta \left( \beta -1/2\right) \textrm{tr}\left[  
\tilde{R}^{0\,2}O_1
\bar{\eta}-\left( \beta \leftrightarrow -\beta \right) \right]\,,
\\
\label{delta3cuenta}
\Delta_{3}&=& -\frac{m^{2}}{2\pi ^{2}}\log (m)\overset{+\infty }{
\underset{1/2}{\int }}d\beta \left( \beta -1/2\right) \textrm{tr}\left[ 
\tilde{R}^{0\,2}{\bf 1}\gamma ^{0}\bar{\eta}\tilde{R}^{0}O_2 \bar{\eta}-\left( \beta \leftrightarrow -\beta
\right) \right]\,.
\end{eqnarray}
where $O_1(x,y)=\left( x-y\right) ^{\mu }\gamma _{\mu }\gamma ^{0}$ and $O_2(x,y)=\log
\left\vert x-y\right\vert\gamma ^{0}$. 

$\Delta_1$ is readily evaluated since the relevant calculation is the same as above,
\begin{equation}
\Delta_1=\frac{\log(2)-\gamma_E}{3}\, \,m^2 \,\log(m)L_+\, L_- \,.
\end{equation}  
The contribution of $C_2$ to this order is obtained following the same steps as for the contribution $S_{2,2}$. This is, using (\ref{teoremagen}) to write the trace in the chiral representation and null coordinates, then using the spectral decomposition of the massless resolvent, and finally, solving the sums and integrals with the help of the formulas in the Appendix. In this case, we have to use (\ref{uya}) because of the explicit dependence of the correlator contribution in the coordinates. 
After some algebra we get an expression in terms of a one-dimensional integral over $V_\pm$,
\begin{equation}
\Delta _{2}=\frac{m^{2}}{2}\log (m)\left[ 
\underset{V_{-}}{\int }du_{-}u_{+}\left( u_{-}\right) \tanh \left(\frac{z_-(u_-)}{2}\right) +\underset{V_{+}}{\int }du_{+}u_{-}\left( u_{+}\right) \tanh \left(\frac{z_+(u_+)}{2}\right)
\right] \,,\label{delta2}
\end{equation}
where we have defined
\begin{equation}
z_\pm(u_\pm)=\log\left(-\frac{\prod_i(u_\pm-a_\pm^i)}{\prod_i(u_\pm-b_\pm^i)}\right)\,.
\end{equation}
For the special case of a set lying in a single spatial hyperplane with total length $L=L_+=L_-$, and choosing the straight line curve, (\ref{delta2}) simplifies to
$\Delta_2=\frac{1}{6} \, \,L^2 \,m^2\,\log(m)$, which is the result presented in \cite{fermion1}.

Finally, the calculation of $\Delta_3$ starts as above, from (\ref{delta3cuenta}), passing to null coordinates through (\ref{teoremagen}), and expanding the resolvent in its spectral decomposition in terms of the eigenvectors (\ref{tisei}). At this point, it is convenient in this case to further write the eigenvectors explicitly using (\ref{hfhf}). The sums over the discrete indices of the eigenvectors are done using (\ref{typy}). Then, one is left with an integral over the eigenvalues $s$ and the resolvent parameter $\beta$. These can be done (preferably in this order), leading to a delta function in the variable $z$. The final result is
\begin{equation}\label{delta3}
\Delta _{3}=-2m^{2}\log(m)\underset{V_{-}}{\int }du_{-}^{y}\underset{V_{+}}{
\int }du_{+}^{x}\log \left\vert x-y\right\vert \delta \left( z\left(
u_{-}^{y}\right) -z\left( u_{+}^{x}\right) \right) \,.
\end{equation}

It has to be noted that the results (\ref{delta2}) and (\ref{delta3}) depend on the chosen curve within the equivalence class of curves having the same domain of dependence as $V$.   The choice of the curve changes for example the function $u_+(u_-)$ in (\ref{delta2}). We know the entropy is independent of this choice, but this is reflected in that the whole contribution $S_{2,1}$ is curve-independent, while $\Delta_2$ and $\Delta_3$ separately are not. We have checked this numerically in several examples. In particular it is possible to choose the curve formed by the null future (or past) horizon of the domain of dependence of $V$, which simplifies some calculations. 
 
For a single interval of length $L$ the integrals can be evaluated analytically and the series reads
\begin{eqnarray}
S(L)&=&\frac{1}{3}\log(L/\epsilon)-\frac{m^2}{6}  \log^2(m) L^2+\frac{\log(2)-\gamma_E+5/6}{3}\, \,m^2 \,\log(m)L^2\nonumber \\ &&\hspace{5.5cm}-\frac{1}{3} \log(L) \, \log(m)\,  m^2 L^2+{\cal O}(m^2 \log^0(m))\,.\label{unok}
\end{eqnarray}
The entropy for a single interval can also be expressed in terms of a solution of a Painlev\'e ordinary differential equation \cite{fermion2}. The expansion for small mass (\ref{unok}) coincides with the one given by this differential equation.  

\begin{figure}[t]
\centering
\leavevmode
\epsfysize=2.cm
\epsfbox{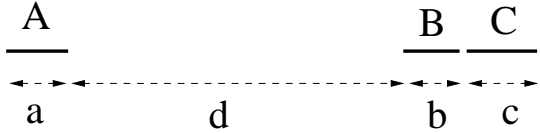}
\bigskip
\caption{Configuration of two adjacent intervals $B$ and $C$ separated from the interval $A$ by a large distance $d$. All three intervals are collinear.  }
\label{hory}
\end{figure}

Of course, for more than one interval the result does not depend only on the total lengths $L_\pm=\sum (b_i^\pm-a_i^\pm)$. In particular, the contributions $\Delta_2$ and $\Delta_3$ to $S_{2,1}$ give place to a non extensive mutual information (see Section 4.2 below). In order to see this analytically in a special limit, consider three collinear intervals $A$, $B$ and $C$, of lengths $a$, $b$ and $c$ respectively. Take the distance from $a$ and $b$ to be $d\gg a,b,c$, and $C$ adjacent to $B$ on the side opposite to $A$ (see figure (\ref{hory})). In the large $d$ limit we can compute the leading term\footnote{This equation corrects a mistake in the eq. (32) of \cite{extensiva}.}
\begin{eqnarray}\label{exphor}
I(A,B,C)\sim m^2 \log(m)\frac{a }{450d^2}\left\{bc\left[-98a+33(b+c)\right]+120 abc\log(a)-30(2a-b)b^2\log(b)+\right.\nonumber\\ \left.30\left\{c^2(-2a+c)\log(c)+(2a-b-c)(b+c)^2\log(b+c)-bc\left[8a-3(b+c)\right]\log(d)\right\}\right\}\,.
\end{eqnarray}

\section{The regime of large relative boosts}
\begin{figure}
\centering
\leavevmode
\epsfysize=5cm
\epsfbox{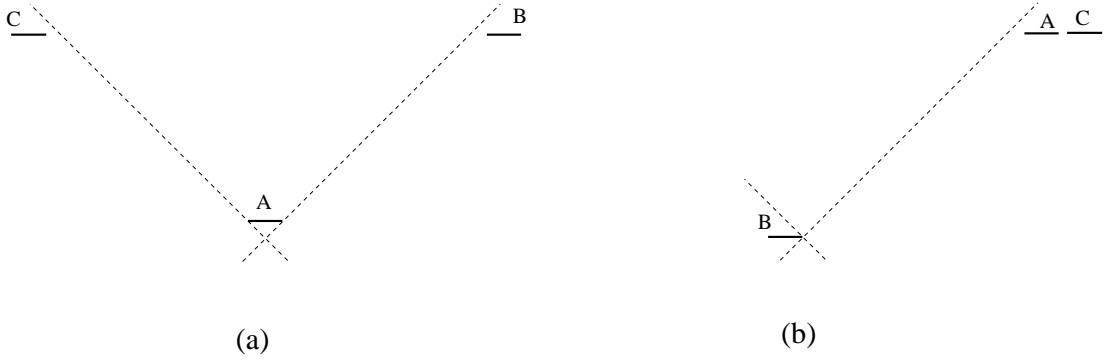}
\bigskip
\caption{Two different configurations of three intervals with a large relative boost between $A$ and $B$ or $C$. The case (a) is extensive in the limit of large relative boosts and fixed sizes $a$, $b$ and $c$ while the case (b) does not show extensivity of the mutual information.}
\label{f44}
\end{figure}

In this section we describe in more detail the behavior of the mutual information in different situations which have in common the presence of large relative boosts between the involved regions. 
Specifically, we focus on the geometric configurations shown in figures (\ref{f33}) and (\ref{f44}). 

\subsection{Two boosted intervals}
We consider two sets, which for simplicity we take to be single intervals of lengths $a$ and $b$, separated by a distance $d$. The separating interval $d$ can be positioned on the $x^1$ axis without loss of generality. In order to completely fix the configuration we have to specify two more parameters, the hyperbolic angles $\alpha$ and $\beta$, determining the relative boosts of $A$ and $B$ with respect to the separating interval. Thus, the size of the projections to the null axis are $a_+=a e^{\alpha}$, $a_-=a e^{-\alpha}$, $b_+=b e^{\beta}$ and $b_-=b e^{-\beta}$. We are interested in the limit of $I(A,B)$ when $a$ and $b$ approach zero. It is easy to see that if $\alpha$ or $\beta$ (or both) are kept bounded while $a$ and $b$ tend to zero then $I(A,B)$ also tends to zero. This is a natural result, since it is expected that for a fixed distance $d$ the mutual information should vanish with the progressive elimination of degrees of freedom in $A$ and $B$. However, if we take the limit of zero size, but at the same time increase the modulus of the boost parameters $\alpha$ and $\beta$ such that the null coordinate projections of $A$ and $B$ are kept finite, then the mutual information takes a finite limit value. 

\begin{figure}[t]
\centering
\leavevmode
\epsfysize=5cm
\epsfbox{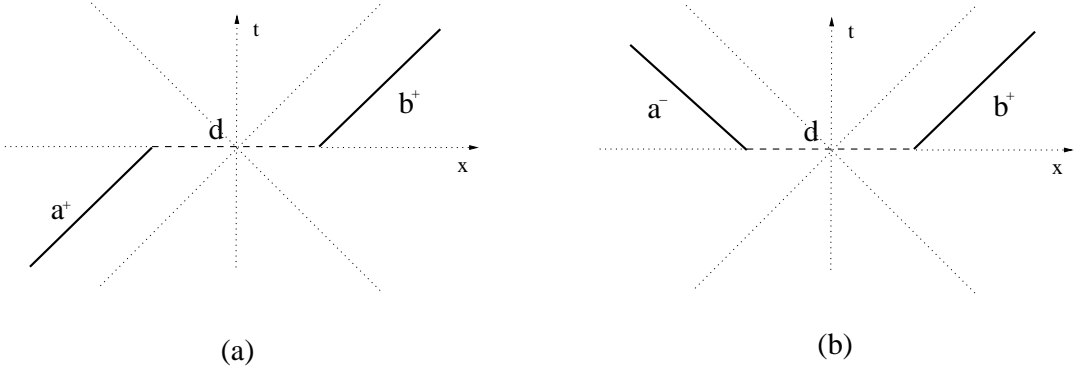}
\bigskip
\caption{Two different limits of vanishing length and divergent relative boost for the intervals $A$ and $B$.  }
\label{f33}
\end{figure}

Two different cases have to be considered. These are shown in figure (\ref{f33}). First the limit in which $\alpha\rightarrow \infty$ and $\beta\rightarrow \infty$, while $a,b\rightarrow 0$, in such a way that $a_+=a e^{\alpha}$ and $b_+=b e^{\beta}$ are finite. In this limit we have $a_-=b_-=0$ (figure(\ref{f33})a). The case of vanishing $a_+$ and $b_+$ ($\alpha, \beta \rightarrow -\infty$) is analogous. The massless contribution to the entropy does not vanish in this case, and we have 
\begin{equation}
I(A,B)\sim\frac{1}{6}\log\left(\frac{(b_++d_+)(a_++d_+)}{d_+ (a_++b_++d_+)}\right)\,.
\end{equation}
Only the cross ratio of the $u^+$ projections matter to this massless contribution. 

Two different ingredients seem to be necessary for an explanation of this result. First, the quantum field theory contains infinite many d.o.f. in any finite volume region of any size. This is important, otherwise the number of d.o.f. would vanish in the zero volume limit\footnote{Infinitely many d.o.f. in a finite volume is also related to the fact that $I(A,B)$ diverges when $A$ and $B$ come into contact for generic $A$ and $B$.}. Here, it also seems important that in the massless limit the theory decomposes into the two chiralities and is conformally invariant. Hence, the plus chirality acts as a one dimensional translational invariant system independent of the minus chirality, and the finite result for $I(A,B)$ is a consequence of the finite values of $a_+$, $b_+$ and $d_+$. The physical size of the $a$ and $b$ intervals is of course zero in the limit, but in a conformal theory there is the same amount of shared information in any rescaled geometric configuration. Thus, the explanation in terms of the conformal limit is that $a$, $b$ and $d$ tend to zero, but only the ratios of their projections in the $u_+$ axis are relevant.   

However, such an explanation combining chiral decomposition with invariance under scaling is helpless in our next example. The second case we want to consider is a limit $a_+=0$, $b_-=0$, $a_-$ and $b_+$  finite (see figure (\ref{f33})b). In this case the conformal contribution vanishes, and the leading log term is given by $S_{2,2}$,
\begin{equation}
I(A,B)\sim  \frac{m^2}{6}  \log^2(m) \,a_+ \,b_-\,.\label{corto}
\end{equation}
Again we have a non zero mutual information for zero volume regions. This case is perhaps more surprising than the previous one since no interpretation as entanglement between small but nearby regions seems possible. It suggests that amplification of entanglement by the large relative boost is a genuine effect that can lead to finite shared information for vanishing small regions separated by a finite, fixed distance. 

In order to clarify the role of the massless limit in this phenomenon we  can look at a different limit for the two boosted intervals. This is a large separating distance limit, when  $d\gg a_-,a_+, b_-,b_+$ and $a_-,a_+,b_-,b_+ \,\times  m\ll 1$, but not necessarily $d m\ll 1$. Thus, at leading order one uses the massless correlator inside the intervals, and considers the correlator between points on $A$ and $B$ as a constant. The calculation then involves the same tools used in the previous Section. The main steps are explained in \cite{fermion1}, and we do not repeat this calculation here. We have for two intervals in the limit of large separation    
\begin{equation}
I(A,B)\sim \frac{1}{6} \,m^2 \,(a_+ b_-+ a_- b_+) \,K_0^2(m d)+\frac{1}{6} \,m^2 \,(a_+ b_++ a_- b_-)  K_1^2(m d) \,.  \label{hghg}
\end{equation}
The first term of the right hand side corresponds to the continuation of (\ref{corto}) to a large separating distance configuration, while the second term corresponds to the continuation of (\ref{miro}). 

Formula (\ref{hghg}) clarifies two issues. First, it is clear that the phenomenon of finite mutual information between null surfaces regions extends out of the conformal limit, for both configurations in the figure (\ref{f33}). Second, since the configurations include large dimensionless boost parameters, one could wonder if the perturbative series is behaving correctly in (\ref{corto}), and if it converges. This was expected, because the large boost parameters appear in the series only in terms of the distances involved and their ratios, which are all finite and bounded. The formula  (\ref{hghg}) extending the result (\ref{corto}) to a different situation, clarifies this expectation is correct.

This behavior of the mutual information for null surfaces is connected with related properties for the correlation functions. The mutual information is a bound for correlations 
\cite{vers},
\begin{equation}
I(A,B)\ge \frac{1}{2}\frac{\left(\langle {\cal O}_A{\cal O}_B\rangle-\langle{\cal O}_A\rangle\langle{\cal O}_B\rangle\right)^2}{\left\|{\cal O}_A\right\|^2\left\|{\cal O}_B\right\|^2}\,,\label{versi}
\end{equation}
where ${\cal O}_A$ and ${\cal O}_B$ are normal operators in $A$ and $B$, and $\left\|{\cal O}\right\|$ is the norm of the operator ${\cal O}$ (the greatest eigenvalue modulus). 
For the fermion field consider the hermitian operators\footnote{An objection is that the operator ${\cal O}_A$ (\ref{op}), being fermionic, is not really localized in $A$, since it does not commute with ${\cal O}_B$ for example. However, the application of (\ref{versi}) is easily generalized to the fermionic case, and it does only require the expectation value of these operators to be given by the corresponding traces involving the density matrices.} \\
\begin{equation}
{\cal O}_A=\int_A ds \, (\alpha^\dagger(s) \Psi(s)+\Psi^\dagger(s) \alpha(s))\,, \hspace{1cm} {\cal O}_B=\int_B ds \, (\beta^\dagger(s) \Psi(s)+\Psi^\dagger(s) \beta(s))\,,\label{op}
\end{equation}
where $\alpha(x)=0$ for $x\notin A$, and $\beta(x)=0$ for $x\notin B$. 
Since the square of these operators are c-numbers, we have for the norms
\begin{equation}
\left\|{\cal O}_A\right\|^2=({\cal O}_A)^2=\int_A ds\, \alpha^\dagger(s)(\bar{\eta}(s))^{-1}\alpha(s)\,,
\end{equation}
and analogously for ${\cal O}_B$.
We also have, $\langle {\cal O}_{A}\rangle=\langle {\cal O}_{B}\rangle=0$ and 
\begin{equation}
\langle {\cal O}_A{\cal O}_B\rangle=\int ds_1\, ds_2\, \left(\alpha^\dagger(s_1)C(s_1,s_2)\beta(s_2)-\beta^\dagger(s_1)C(s_1,s_2)\alpha(s_2)\right)\,.
\end{equation}
Then, using the Cauchy-Schwartz inequality, we have for the right hand side of (\ref{versi}) 
\begin{equation}
\frac{\left(\langle {\cal O}_A{\cal O}_B\rangle-\langle{\cal O}_A\rangle\langle{\cal O}_B\rangle\right)^2}{\left\|{\cal O}_A\right\|^2\left\|{\cal O}_B\right\|^2}\le   \,a \,b \,\left\|\bar{\eta}  \right\|_A \left\|\bar{\eta}\right\|_B\,\left\|C\right\|_{A,B}^2\,,\label{labe}
\end{equation}
where $a$ and $b$ are the lengths of $A$ and $B$, $\left\|\bar{\eta}  \right\|_V$ is the maximum of the norm of $\bar{\eta}(x)$ for $x\in V$, and  $\left\|C\right\|_{A,B}=
\max_{x\in A,\, y\in B}\left\|C(x,y)\right\|$ is the maximum of the absolute value of the eigenvalues of correlator between $A$ and $B$. 
Then, if the slope of the curve is bounded, $\left\|\bar{\eta}  \right\|_A$ and $\left\|\bar{\eta}  \right\|_B$ are bounded according to (\ref{etaeta}). There is no possibility for (\ref{labe}) but to vanish when $a\rightarrow 0$ or $b\rightarrow 0$. This is consistent with $I(A,B)\rightarrow 0$ in these cases. However, if $a,b\rightarrow 0$ but $a\, b\, \left\|\bar{\eta}\right\|_A\,\left\|\bar{\eta}\right\|_B$ remains finite (or diverges) in the limit, one can produce a non zero lower bound for the mutual information through (\ref{versi}). This is clearly what happens when one takes $a, b\rightarrow 0$, but keeping their projections $a^\pm, b^\pm$ finite. Hence, in the language of operators our observation is that boosts allow for a large relative enhancing of operator correlations as compared to operator norm.

We have chosen to put $d$ horizontal in figure (\ref{f33}) without loss of generality. In the case of $d\rightarrow 0$,  the mutual information is either divergent or ambiguous. This last possibility is exemplified by the case $a^-,b^-,d^-\rightarrow 0$, i.e. $A$, $B$ and $D$ all tend to lie in a null surface. The mutual information (\ref{miro}) depends in this case on the ratios of $a^-$, $b^-$ and $d^-$ as they tend to zero. 

\subsection{Extensivity of the mutual information}
Now we turn attention to the analysis of the extensivity properties of the mutual information. 

The first thing to note is that in general the relative importance of the non extensivity with respect to the mutual information does not necessarily decrease with decreasing correlations. For example, in the configuration of three coplanar intervals of figure (\ref{hory}), the mutual information is dominated by the massless contribution (\ref{miro}) (we take $a,b,c\ll d\ll m^{-1}$), while $I(A,B,C)$ is given by the leading massive term (\ref{exphor}).  Then we have,
\begin{equation}
\frac{I(A,B,C)}{I(A,BC)}\sim \frac{b c  (-8a+3(b+c))\log(d)}{5  (b+c)} \,m^2 \log(m)\,.\label{iu}
\end{equation}
This is increasing with the distance $d$, at least in the range $d\ll m^{-1}$.

A partial understanding for this non-extensivity for the configuration of the figure (\ref{hory}) follows from the geometry. The typical distances $\textrm{dis}(A,B)\sim \textrm{dis}(A,C)\sim d\gg 1$ while $\textrm{dis}(B,C)\in (0,b+c)$. This is a typical Euclidean regime, where the triangle inequalities are satisfied. In this case the geometry cannot enforce extensivity for the mutual information. This is because while $A$ is correlated similarly with $B$ and $C$, the correlation between $B$ and $C$ is higher or similar to the one they have with $A$. Thus, $A$ is not entangled independently with $B$ and $C$, but rather there is an important part of the correlations which are truly tripartite, as shown by the actual calculation of $I(A,B,C)$ (see (\ref{iu})).

\begin{figure}
\centering
\leavevmode
\epsfysize=6cm
\epsfbox{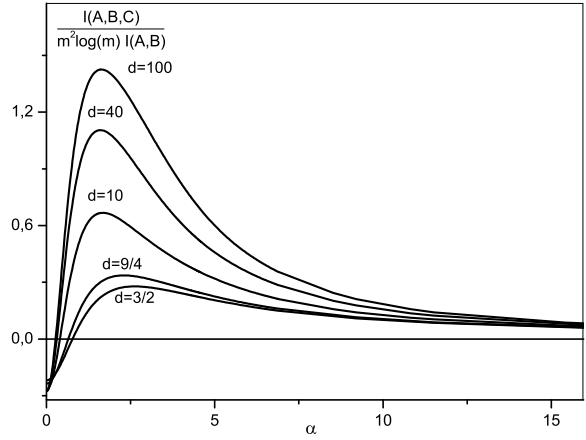}
\bigskip
\caption{Relative extensivity of the mutual information for the configuration of figure (\ref{f44})a. The interval sizes are $a=b=c=1$.} 
\label{fesi}
\end{figure}

In this same line of thought, one could expect that if there is a condition enforcing extensivity which is general enough to be useful to the applications on black hole physics, the configuration has to be largely non-euclidean, with great violations of the triangle inequalities in the distances. 

A possibility is to search for situations where $I(B,C)\ll I(A,B), I(A,C)$. This is typically the case of the Hawking radiation, with large and uncorrelated asymptotic regions (represented by $B$ and $C$) which are however correlated with the black hole (represented by $A$). In this case the state $\rho_{BC}$ is well approximated by a product state. This is still not equivalent to extensivity since in this case ($I(B,C)\sim 0$) one has only a one sided inequality,
\begin{equation}
\frac{I(A,B,C)}{I(A,B)}= \frac{I(A,B)+I(B,C)-I(B,AC)}{I(A,B)}< \frac{I(B,C)}{I(A,B)}\sim 0 \,.
\end{equation}     
Here we have used the monotonicity property of the mutual information, $I(B,AC)>I(B,A)$. In the context of quantum entanglement measures a relation like this one, which corresponds to $E(A,B,C)=E(A,B)+E(A,C)-E(A,BC)\le 0$ for the entanglement measure $E$, has been called monogamy of entanglement \cite{monogamy}. Some entanglement measures (for example the squashed entanglement \cite{ee}) always satisfy this inequality. This is not the case of the mutual information, which also measures classical correlations, and in general is non-monogamous, excepting special situations as the one we are focusing here.

\begin{figure}
\centering
\leavevmode
\epsfysize=6cm
\epsfbox{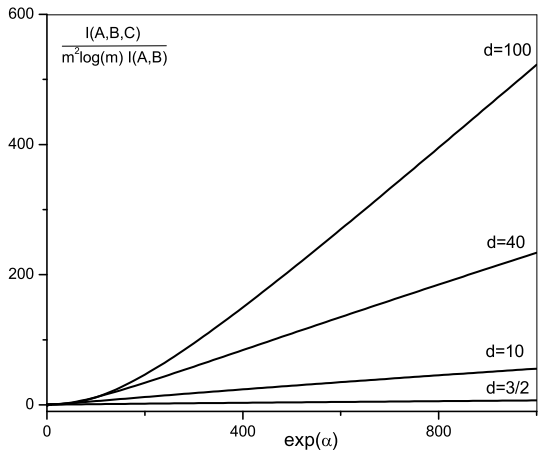}
\bigskip
\caption{Relative extensivity of the mutual information for the configuration of figure (\ref{f44})b. The interval sizes are $a=b=c=1$.}
\label{fese}
\end{figure}

In order to have $I(B,C)\ll I(A,B), I(A,C)$ we need in general that $\textrm{dis}(A,B)\sim \textrm{dis}(A,C)\ll \textrm{dis}(B,C)$, a deeply Lorenzian regime. The figure (\ref{f44})a shows one such a configuration. We again have three intervals $A$, $B$, $C$ parallel to the $x^1$ axes but now there is a high relative boost between $A$ and $B$, and $A$ and $C$, with opposite signs. Specifically, we write the interval between the right-most point in $C$ and the left-most point in $A$ as $d (-\sinh(\alpha),\cosh(\alpha))$, and the interval between the right-most point in $A$ and the left-most point in $B$ to be $d (\sinh(\alpha),\cosh(\alpha))$. We are interested in the behavior of the non-extensivity as a function of $\alpha$. The typical distances are $\textrm{dis}(A,B)=\textrm{dis}(A,C)\sim d$, and $\textrm{dis}(B,C)\sim e^{\alpha/2} d$ for large $\alpha$. An example with $a=b=c=1$ and several values of $d$ is shown in figure (\ref{fesi}). 
  The relevant entropy contributions to $I(A,B,C)$, computed by eqs. (\ref{delta2}) and (\ref{delta3}) are given by integrals in one variable. We evaluate them numerically.
We see the relative tripartite information starts negative for small $\alpha$. Then, with increasing boost it changes sign and peaks with a positive value. For large boost parameter $\alpha$ it decreases asymptotically like $1/\alpha$. Hence, our expectations are confirmed in this simple example where the mutual information is extensive in the limit of large relative boosts.

It is interesting to note that well in the large boost situation the tripartite information is positive, violating the strict monogamy relation $I(A,B,C)<0$. This is good, since when this relation holds we necessarily have extensivity in the limit of large $\alpha$ because  $0<I(A,B,C)/I(A,B)<I(B,C)/I(A,B)\rightarrow 0$.  
Note also that the case of fixed $\alpha$ and increasing $d$ does not lead to $I(A,B,C)/I(A,B)\rightarrow 0$, similarly to what happens for the collinear case (\ref{iu}). This is expected in accordance with the fact that the distances $\textrm{dis}(A,B)$, $\textrm{dis}(A,C)$, and $\textrm{dis}(B,C)$ all increase proportional to $d$ in this case, and there are no large violations of the triangle inequalities. 

 However, a connection between triangle inequalities and extensivity of information is not the whole story, and a further consideration of the relative boosts between the regions is necessary. Consider the geometry of figure (\ref{f44})b. In this case even if the configuration enhances the correlations of $A$ with $B$ and $C$ over the one between $B$ and $C$, by approaching $A$ and $B$ to a common null line, this is done in a non fully satisfactory way. We again take $A$, $B$ and $C$ parallel to the $x^1$ axis, the distance between $A$ and $B$ equal to $d$, and the interval separating the right-most point of $B$ to the left-most point in $A$ to be $d(\sinh(\alpha),\cosh(\alpha))$. 
The typical distances are $\textrm{dis}(A,B)\sim d$, $\textrm{dis}(A,C)\in (0,a+c)$ and $\textrm{dis}(B,C)\sim e^{\alpha/2} \sqrt{2 a d}$ for large $\alpha$.  
 The figure (\ref{fese}) shows an example with $a=b=c=1$ and several values of the separating distance $d$. The quantity of interest is the relative importance of the non-extensivity, measured by $I(A,B,C)/(I(A,B)m^2 \log(m))$. The mutual information $I(A,B)$ is dominated by the massless contribution, which in the large $\alpha$ limit behaves linearly $I(A,B)\simeq \alpha$. However, the tripartite information is linear in the boost factor $\exp(\alpha)$, as shown in figure (\ref{fese}). Therefore the relative non-extensivity grows rapidly with the boost. 
 
Therefore, in this case, even if the triangle inequalities are largely violated we still do not have extensivity. The reason can be traced back to the fact that here large $\textrm{dis}(B,C)$ does not mean small $I(B,C)$. This is the same phenomenon we discussed in the last section. The distance between $B$ and $C$ increases as $e^{\frac{\alpha}{2}}$ but the relative boost of $B$ and $C$ with respect to the separating interval is also proportional to this factor. In fact, the massless contribution gives a non vanishing $I(B,C)$ in the limit of large $\alpha$.  In order to have asymptotic extensivity a bound on the relative boosts seems to be necessary.

\section{Concluding remarks}
We have obtained the entanglement entropy for a free fermion in two dimensions for a completely general  relativistic region. This has been done in a small mass approximation.  

We found a large amplification of shared information by boosts. This gives a non zero mutual information between two sets of arbitrary small size at a fixed minimum distance between each other, provided the relative boost diverges in the limit of small size. A partial understanding of this behavior in terms of correlators was provided. In more dimensions the analog effect is a non zero mutual information between null surface regions.

This effect can also be understood in terms of the operator product expansion (OPE) of certain operators. In two dimensions the traces of integer powers of the reduced density matrix,  $\textrm{tr}(\rho_V^n)=e^{-(n-1)S_n(V)}$, where $S_n(V)$ is the Renyi entropy of index $n$ corresponding to $V$, can be written in terms of operator correlators \cite{mio,twist}. We have 
\begin{equation}
e^{-(n-1) S_n(l_1,r_1,...,l_k,r_k)}=\langle 0 \vert \tilde{\Phi}_n(l_1) \Phi_n(r_1)... \tilde{\Phi}_n(l_k)\Phi_n(r_k) \vert 0\rangle\,,
\end{equation}
where $V=(l_1,r_1,...,l_k,r_k)$ contains $k$ intervals, and $\tilde\Phi_n(x)$ is the CPT conjugate of the field $\Phi_n(-x)$. In the Euclidean formulation the fields $\Phi_n(x)$ are a special class of twisting operators \cite{twist}. The entropy is a limit of the Renyi entropies for index $n\rightarrow 1$, and can be obtained by analytic continuation in the variable $n$. Then, it is expected (and often found) that the Renyi entropies and the entropy have a similar behavior as functions of the coordinates. In this representation the limit of small size of the $j^{\textrm{th}}$ interval corresponds to the limit of the correlator in which there is a product $\tilde{\Phi}_n(l_j) \Phi_n(r_j)$ for $l_j\rightarrow r_j$. In this limit we are able to use the OPE \cite{twist}
\begin{equation}
\tilde{\Phi}_n(-x/2) \Phi_n(x/2)\simeq c \vert x \vert^{-\frac{n-1/n}{6}} \, {\bf 1}\,,\label{hhhh}
\end{equation}
which is proportional to the identity operator to leading order.
Hence, the limit of the Renyi mutual information $I_n(A,B)=S_n(A)+S_n(B)-S_n(AB)$ for sets containing an interval with vanishing small length emerges naturally from (\ref{hhhh}).
 This reads simply
\begin{equation}
\lim_{l^A\rightarrow r^A} I_n((l^A,r^A),(l_1^B,r_1^B,...,l_q^B,r_q^B))=0\,,
\end{equation}
and 
\begin{eqnarray}
&&\lim_{l_j^A\rightarrow r_j^A} I_n((l_1^A,r_1^A,...,l_j^A, r_j^A,...,l_k^A,r_k^A),(l_1^B,r_1^B,...,l_q^B,r_q^B))=\\&&\hspace{6cm}I_n((l_1^A,r_1^A,...,\check{l}_j^A, \check{r}_j^A,...,l_k^A,r_k^A),(l_1^B,r_1^B,...,l_q^B,r_q^B))\,,\nonumber
\end{eqnarray}
where in the right hand side the marked points on the Renyi mutual information argument are absent, and thus the first set has only $k-1$ intervals. That is, the vanishing small size interval just disappear from the argument of the mutual information.  

However, if we take the limit $l^A\rightarrow r^A$ by approaching a null line (while leaving all other points fixed) the OPE has a very different content. This is because boost and distance are scaled proportionally to each other in this limit, and the series involves an infinite number of operators of increasing dimension and spin even at leading order \cite{ope}. The impossibility of replacing the product of the two approaching operators by a finite combination of operators at some given point is clearly seen for example in figure (\ref{f33}). Even if the two end points $l_A$ and $r_A$ of $A$ are at small distance from each other, their distances $\textrm{dis}(l_A,x)$ and $\textrm{dis}(r_A,x)$ from a point $x\in B$ are very different.    

Based on the lessons learned so far from this model, we are tempted to conjecture that the mutual information is extensive, $I(A,B,C)/I(A,B)\rightarrow 0$, in the limit of large violation of the triangle inequalities $\textrm{dis}(A,B)\sim \textrm{dis}(A,C)\ll \textrm{dis}(B,C)$, and when the relative boosts involved in the configuration of $BC$ are kept bounded. This is the appropriate geometric scenario for analyzing the information shared by a black hole with the radiation region. Indeed, we are also tempted to suggest that this may be a general effect, leading to information extensivity in the Hawking radiation.  

A proof of this conjecture in Minkowski space may involve the analysis of the clustering properties of the operators in this particular geometrical limit, and a more complete understanding of the relation between the mutual information and the correlators than the one provided by (\ref{versi}).  This configuration corresponds to the OPE in a limit involving several momentum becoming large simultaneously. This is reminiscent to \cite{tho}, where the information content of Hawking radiation was linked to the understanding of virtual particle collisions in the limit of very large centre of mass energies. 

\section*{Appendix: Spectral resolution of the kernel $(x-y)^{-1}$}

In this appendix we display the spectral decomposition of the kernel $(x-y)^{-1}$ (with principal value regularization) in a generic multi-interval set $V=(a_1,b_1)\cup...\cup(a_n,b_n)\subseteq R$, and collect the properties of the eigenvectors, which are relevant to the calculations in the main text. Most of these properties appear in \cite{fermion1} (see also \cite{mush}).    

The eigenvectors $\Psi^k_s$, 
\begin{equation}
\int_V dy \, \frac{1}{x-y} \Psi^k_s(y)=\lambda_s \Psi^k_s(x)\,,\hspace{1cm} k=1,..,n,
\end{equation}
corresponding to the eigenvalue 
 \begin{equation}
\lambda_s =i \pi \tanh(\pi s)\in (-i\pi,i\pi) \label{eigen}\,
\end{equation}
for any $s\in R$, are $n$-fold degenerate, where $n$ is the number of intervals. 
An explicit expression for an orthonormal basis of eigenvectors $\Psi^k_s(x)$ reads
 \begin{equation}
\Psi_s^k(x)=\frac{ e^{(-i\, s+1/2)\, z(x)}}{N_k\, (x-a_k)}\, \,,\label{hfhf}
\end{equation}
where 
 \begin{eqnarray}
z(x)&=&\log\left(-\frac{\prod_{i=1}^n (x-a_i)}{\prod_{i=1}^n (x-b_i)}\right) \,, \label{bfbf}\\
N_k&=&\sqrt{\pi} \left(\sum_{j=1}^n  \frac{\prod_{l\neq k} (b_j-a_l)}{(b_j-a_k) \prod_{l\neq j} (b_j-b_l)}-\frac{\prod_{j\neq k} (a_k-a_j)}{\prod_{j=1}^n (a_k-b_j)}\right)^{\frac{1}{2}}\,.
\end{eqnarray}

The eigenvectors satisfy the orthonormality relations,
\begin{eqnarray}
\int_V dx \, \Psi^{k*}_s(x) \Psi^{k^\prime}_{s^\prime}(x)&=&\delta_{k,k^\prime} \, \delta(s-s^{\prime})\,, 
\\
\sum_{k=1}^n \int_{-\infty}^\infty ds\, \Psi^{k*}_s(x) \Psi^{k}_{s}(y)&=&\delta(x-y)\,.
\end{eqnarray}
Besides we have the integrals
\begin{eqnarray}
\int_V dx\, \Psi^{k}_s(x)&=&(-1)^{n+1} \pi \, \textrm{sech}(\pi s) N_k^{-1}\,,\label{nos}\\
\textrm{Im}\int_V dx\,x\, \Psi^{k}_s(x)&=& (-1)^{n}  s \pi \,\textrm{sech}(\pi s) N_k^{-1}    L\,,\label{uya}
\end{eqnarray}
where $L=\sum_1^n (b_i-a_i)$ is the sum of the $n$ interval lengths. 

We also employ the useful algebraic identities 
\begin{eqnarray}
\sum_{i=1}^n N_i^{-2}&=&\frac{\sum_{i=1}^n (b_i-a_i)}{2\pi}=\frac{L}{2\pi}\,,\label{nuno}
\\
\sum_{k=1}^n N_k^{-2} (x-a_k)^{-1}&=&\frac{1}{2\pi} (1+e^{-z})\,.\label{typy}
\end{eqnarray}

\section*{Acknowledgments}
This work was partially supported by CONICET, Universidad Nacional de Cuyo, and CNEA, Argentina.
    

\begin{thebibliography}{99}

\bibitem{bh}
L.~ Bombelli, R.~ K. Koul, J.~ Lee, and R.~ D.~Sorkin,  Phys. \ Rev. \ D {\bf 34}, 373383 (1986);
%\cite{Callan:1994py}
  C.~G.~.~Callan and F.~Wilczek,
  %``On geometric entropy,''
  Phys.\ Lett.\  B {\bf 333}, 55 (1994)
  [arXiv:hep-th/9401072].
  %%CITATION = PHLTA,B333,55;%%

\bibitem{cteo}
H.~Casini and M.~Huerta,
%``A finite entanglement entropy and the c-theorem,''
Phys.\ Lett.\ B {\bf 600}, 142 (2004)
[arXiv:hep-th/0405111]. 
%%CITATION = HEP-TH 0405111;%%

\bibitem{rfl}
%\cite{Myers:2010xs}
  R.~C.~Myers and A.~Sinha,
  %``Seeing a c-theorem with holography,''
  Phys.\ Rev.\  D {\bf 82}, 046006 (2010)
  [arXiv:1006.1263];
  %%CITATION = PHRVA,D82,046006;%%
  %\cite{Myers:2010tj}
  R.~C.~Myers, A.~Sinha,
  %``Holographic c-theorems in arbitrary dimensions,''
  JHEP {\bf 1101}, 125 (2011) 
  [arXiv:1011.5819].

\bibitem{top}
 %\cite{Kitaev:2005dm}
  A.~Kitaev and J.~Preskill,
  %``Topological entanglement entropy,''
  Phys.\ Rev.\ Lett.\  {\bf 96}, 110404 (2006)
  [arXiv:hep-th/0510092].
  %%CITATION = HEP-TH 0510092;%%
  
\bibitem{pt}
See for example G. Vidal, J. I. Latorre, E. Rico, A. Kitaev, Phys. Rev. Lett. {\bf 90}, 227902 (2003) [arXiv:quant-ph/0211074].
 
\bibitem{con}
  %\cite{Klebanov:2007ws}
  I.~R.~Klebanov, D.~Kutasov and A.~Murugan,
  %``Entanglement as a Probe of Confinement,''
  Nucl.\ Phys.\  B {\bf 796}, 274 (2008)
  [arXiv:0709.2140].
  %%CITATION = NUPHA,B796,274;%%
  
\bibitem{ryu}
  S.~Ryu and T.~Takayanagi,
  %``Holographic derivation of entanglement entropy from AdS/CFT,''
  Phys.\ Rev.\ Lett.\  {\bf 96}, 181602 (2006)
  [arXiv:hep-th/0603001];
  %%CITATION = PRLTA,96,181602;%%
S.~Ryu and T.~Takayanagi,
%``Aspects of holographic entanglement entropy,''
JHEP {\bf 0608}, 045 (2006)
[arXiv:hep-th/0605073];
%%CITATION = JHEPA,0608,045;%%
%\cite{Nishioka:2009un}
  T.~Nishioka, S.~Ryu and T.~Takayanagi,
  %``Holographic Entanglement Entropy: An Overview,''
  J.\ Phys.\ A  {\bf 42}, 504008 (2009)
  [arXiv:0905.0932].
  %%CITATION = JPAGB,A42,504008;%%
 For very recent developments see for example \cite{rfl} and: 
 %\cite{Headrick:2010zt}
  M.~Headrick,
  %``Entanglement Renyi entropies in holographic theories,''
  Phys.\ Rev.\  {\bf D82}, 126010 (2010) 
  [arXiv:1006.0047]; 
%\cite{Casini:2011kv}
  H.~Casini, M.~Huerta, R.~C.~Myers,
  %``Towards a derivation of holographic entanglement entropy,'' 
  [arXiv:1102.0440];
  %\cite{Tonni:2010pv}
  E.~Tonni,
  %``Holographic entanglement entropy: near horizon geometry and disconnected regions,'' 
  [arXiv:1011.0166]; 
%\cite{deBoer:2011wk}
  J.~de Boer, M.~Kulaxizi, A.~Parnachev,
  %``Holographic Entanglement Entropy in Lovelock Gravities,'' 
  [arXiv:1101.5781].

\bibitem{geo}
%\cite{Casini:2003ix}
  H.~Casini,
  %``Geometric entropy, area, and strong subadditivity,''
  Class.\ Quant.\ Grav.\  {\bf 21}, 2351-2378 (2004) 
  [arXiv:hep-th/0312238].

\bibitem{uni}
C. Adami, [arXiv:quant-ph/0405005]; 
A. Peres and D. R. Terno, Rev. Mod. Phys. {\bf 76}, 93 (2004) [arXiv:quant-ph/0212023]. 

\bibitem{review}
H.~Casini and M.~Huerta,
%``Entanglement entropy in free quantum field theory,''
J.\ Phys.\ A  {\bf 42}, 504007 (2009)
[arXiv:0905.2562].

\bibitem{pes}
I.~Peschel, J.\ Phys.\ A: Math.\ Gen. {\bf 36}, L205 (2003) [arXiv:cond-mat/0212631]. 

\bibitem{fermion1}
H.~Casini and M.~Huerta,
%``Reduced density matrix and internal dynamics for multicomponent regions,''
Class.\ Quant.\ Grav.\  {\bf 26}, 185005 (2009)
[arXiv:0903.5284].
%%CITATION = CQGRD,26,185005;%

\bibitem{narn}
%\cite{Narnhofer:2002ic}
  H.~Narnhofer,
  %``Entanglement, split and nuclearity in quantum field theory,''
  Rept.\ Math.\ Phys.\  {\bf 50}, 111 (2002); 
%\cite{Narnhofer:1993fa}
  H.~Narnhofer,
  %``Entropy density for relativistic quantum field theory,''
  Rev.\ Math.\ Phys.\  {\bf 6}, 1127 (1994).

\bibitem{bhh}
%\cite{Casini:2007dk}
  H.~Casini,
  %``Entropy localization and extensivity in the semiclassical black hole evaporation,''
  Phys.\ Rev.\  {\bf D79}, 024015 (2009) 
  [arXiv:0712.0403].

\bibitem{extensiva} 
%\cite{Casini:2008wt}
  H.~Casini, M.~Huerta,
  %``Remarks on the entanglement entropy for disconnected regions,''
  JHEP {\bf 0903}, 048 (2009) 
  [arXiv:0812.1773].

\bibitem{reh}
%\cite{Longo:2009mn}
  R.~Longo, P.~Martinetti, K.~H.~Rehren,
  %``Geometric modular action for disjoint intervals and boundary conformal field theory,''
  Rev.\ Math.\ Phys.\  {\bf 22}, 331-354 (2010) 
  [arXiv:0912.1106].

\bibitem{fermion2}%\cite{Casini:2005rm}
  H.~Casini, C.~D.~Fosco, M.~Huerta,
  %``Entanglement and alpha entropies for a massive Dirac field in two dimensions,''
  J.\ Stat.\ Mech.\  {\bf 0507}, P07007 (2005) 
  [cond-mat/0505563].

\bibitem{wi}
%\cite{Hertzberg:2010uv}
  M.~P.~Hertzberg, F.~Wilczek,
  %``Some Calculable Contributions to Entanglement Entropy,''
  Phys.\ Rev.\ Lett.\  {\bf 106}, 050404 (2011) 
  [arXiv:1007.0993].

\bibitem{mush} The resolvent of the kernel appears in: N. I. Muskhelishvili, {\sl Singular Integral Equations}, Groningen-Holland (1953), chapter 14. 

\bibitem{hol}
 C.~Holzhey, F.~Larsen and F.~Wilczek,
 %``Geometric and renormalized entropy in conformal field theory,''
 Nucl.\ Phys.\ B {\bf 424}, 443 (1994)
 [arXiv:hep-th/9403108].
 %%CITATION = HEP-TH 9403108;%%

\bibitem{cc1}
%\cite{Alba:2011fu}
  V.~Alba, L.~Tagliacozzo, P.~Calabrese,
  %``Entanglement entropy of two disjoint intervals in c=1 theories,''
  [arXiv:1103.3166];
%\cite{Calabrese:2010he}
  P.~Calabrese, J.~Cardy, E.~Tonni,
  %``Entanglement entropy of two disjoint intervals in conformal field theory II,''
  J.\ Stat.\ Mech.\  {\bf 1101}, P01021 (2011)
  [arXiv:1011.5482];
%\cite{Fagotti:2010yr}
  M.~Fagotti, P.~Calabrese,
  %``Entanglement entropy of two disjoint blocks in XY chains,''
  J.\ Stat.\ Mech.\  {\bf 1004}, P04016 (2010)
  [arXiv:1003.1110];
%\cite{Calabrese:2009ez}
  P.~Calabrese, J.~Cardy, E.~Tonni,
  %``Entanglement entropy of two disjoint intervals in conformal field theory,''
  J.\ Stat.\ Mech.\  {\bf 0911}, P11001 (2009)
  [arXiv:0905.2069];
  %\cite{Alba:2009ek}
  V.~Alba, L.~Tagliacozzo, P.~Calabrese,
  %``Entanglement entropy of two disjoint blocks in critical Ising models,''
  [arXiv:0910.0706];
%\cite{Furukawa:2008uk}
  S.~Furukawa, V.~Pasquier, J.~'i.~Shiraishi,
  %``Mutual Information and Compactification Radius in a c=1 Critical Phase in One Dimension,''
  [arXiv:0809.5113].

\bibitem{vers} M.M. Wolf, F. Verstraete, M.B. Hastings, J.I. Cirac, 
   Phys.\ Rev.\ Lett.\ {\bf 100}, 070502 (2008) 
[arXiv:0704.3906].

\bibitem{monogamy} See for example R. Horodecki, P. Horodecki, M. Horodecki and K. Horodecki, Rev. Mod. Phys. {\bf 81}, 865 (2009) [arXiv:quant-ph/0702225].

\bibitem{ee}M. Koashi and A. Winter, Phys. Rev {\bf A 69}, 022309 (2004) [arXiv: quant-ph/0310037].

\bibitem{mio}
%\cite{Casini:2010nn}
  H.~Casini,
  %``Entropy inequalities from reflection positivity,''
  J.\ Stat.\ Mech.\  {\bf 1008}, P08019 (2010)
  [arXiv:1004.4599].
  %%CITATION = JSTAT,1008,P08019;%%

\bibitem{twist}
%\cite{Calabrese:2004eu}
  P.~Calabrese and J.~L.~Cardy,
  %``Entanglement entropy and quantum field theory,''
  J.\ Stat.\ Mech.\  {\bf 0406}, P002 (2004)
  [arXiv:hep-th/0405152];
  %%CITATION = JSTAT,0406,P002;%%
%\cite{Calabrese:2009qy}
  P.~Calabrese and J.~Cardy,
  %``Entanglement entropy and conformal field theory,''
  J.\ Phys.\ A  {\bf 42}, 504005 (2009)
  [arXiv:0905.4013];
  %%CITATION = JPAGB,A42,504005;%%
%\cite{Cardy:2007mb}
  J.~L.~Cardy, O.~A.~Castro-Alvaredo and B.~Doyon,
  %``Form factors of branch-point twist fields in quantum integrable models and
  %entanglement entropy,''
   [arXiv:0706.3384];
  %%CITATION = ARXIV:0706.3384;%%
%\cite{Caraglio:2008pk}
  M.~Caraglio and F.~Gliozzi,
  %``Entanglement Entropy and Twist Fields,''
  JHEP {\bf 0811}, 076 (2008)
  [arXiv:0808.4094].
  %%CITATION = JHEPA,0811,076;%%

\bibitem{ope}
%\cite{Brandt:1970kg}
  R.~A.~Brandt and G.~Preparata,
  %``Operator Product Expansions Near The Light Cone,''
  Nucl.\ Phys.\  B {\bf 27}, 541 (1972).
  %%CITATION = NUPHA,B27,541;%%

\bibitem{tho}
%\cite{'tHooft:1984re}
  G.~'t Hooft,
  %``On The Quantum Structure Of A Black Hole,''
  Nucl.\ Phys.\  B {\bf 256}, 727 (1985); 
  %%CITATION = NUPHA,B256,727;%%
%\cite{Stephens:1993an}
  C.~R.~Stephens, G.~'t Hooft and B.~F.~Whiting,
  %``Black hole evaporation without information loss,''
  Class.\ Quant.\ Grav.\  {\bf 11}, 621 (1994)
  [arXiv:gr-qc/9310006].
  %%CITATION = CQGRD,11,621;%%

\end{thebibliography}
\end{document}